%% file: combine.tex
\documentclass[conference]{IEEEtran}
\IEEEoverridecommandlockouts
\usepackage{cite}
\usepackage{amsmath,amssymb,amsfonts}
\usepackage{algorithmic}
\usepackage{graphicx}
\usepackage{textcomp}
\usepackage{xcolor}
\usepackage{hyperref}
\usepackage[capitalize]{cleveref}
\usepackage{subcaption}
\usepackage{booktabs}
\usepackage{multirow}
\usepackage[table]{xcolor}
\usepackage{hhline}

\def\BibTeX{{\rm B\kern-.05em{\sc i\kern-.025em b}\kern-.08em
    T\kern-.1667em\lower.7ex\hbox{E}\kern-.125emX}}
\begin{document}

\title{TempRetinex: Retinex-Based Unsupervised Enhancement for Low-Light Video Under Diverse Lighting Conditions}

\author{\IEEEauthorblockN{Yini Li,
Louis Forster,
David Bull,
Nantheera Anantrasirichai}


\IEEEauthorblockA{
Visual Information Laboratory, School of Computer Science, University of Bristol, Bristol, UK\\
Email: \{ub24017, louis.forster.2022, Dave.Bull, N.Anantrasirichai\}@bristol.ac.uk}

}

\maketitle

\begin{abstract}
The acquisition of paired low-light video sequences remains challenging due to issues associated with poor temporal consistency, varying illumination characteristics and camera parameters. This has driven significant interest in unsupervised low-light enhancement approaches. In this context, we propose TempRetinex, an unsupervised Retinex-based video enhancement framework exploiting inter-frame correlations. We introduce Brightness Consistency Preprocessing (BCP) that explicitly aligns intensity distributions across exposures. BCP is shown to significantly improve model robustness to diverse lighting scenarios. Moreover, we propose a multiscale temporal consistency-aware loss and an occlusion-aware masking technique to enforce similarity between consecutive frames. We further incorporate a Reverse Inference (RI) strategy to refine temporally unstable frames and a Self-Ensemble (SE) mechanism to boost denoising across diverse textures. Experiments demonstrate that TempRetinex achieves state-of-the-art performance in perceptual quality. Code is available at \url{https://github.com/liyinibristol/TempRetinex}.
\end{abstract}



\section{Introduction}
\label{sec:intro}

Low-light enhancement refers to the improvement of visual quality for content captured under conditions of poor illumination. While increasing camera ISO allows a sensor to capture more photons, it also amplifies sensor noise, making joint color correction and noise suppression essential. This task is important not only for visualization, but also for downstream computer vision tasks \cite{anantrasirichai2022artificial}.

Recent deep learning-based methods have greatly enhanced the performance of low-light enhancement algorithms  \cite{lin2024sunet}, but several challenges remain. A key issue is the scarcity of high-quality paired video datasets, which limits the generalization ability of supervised approaches. Most methods also underutilize temporal information, simply extending single-image processing techniques to process multiple individual frames; thus introducing flickering artifacts. Variations in camera settings (ISO, aperture and exposure) under changing lighting conditions result in heterogeneous video characteristics, further complicating model training and generalization. 

Video-specific unsupervised low-light enhancement methods remain scarce. Existing techniques either process frames independently \cite{Zheng:semantic:2022} or fuse adjacent frames to form the input \cite{10210621}. Zero-TIG \cite{li2025zerotig} made progress by introducing temporal feedback but it lacks explicit temporal constraints. To address these limitations, we introduce \textbf{TempRetinex}, an unsupervised Retinex-based framework featuring a multiscale temporal consistency loss and a self-ensemble mechanism for robust video enhancement. The key innovations include:

\begin{figure*}

  \centering
  \includegraphics[width=0.95\linewidth]{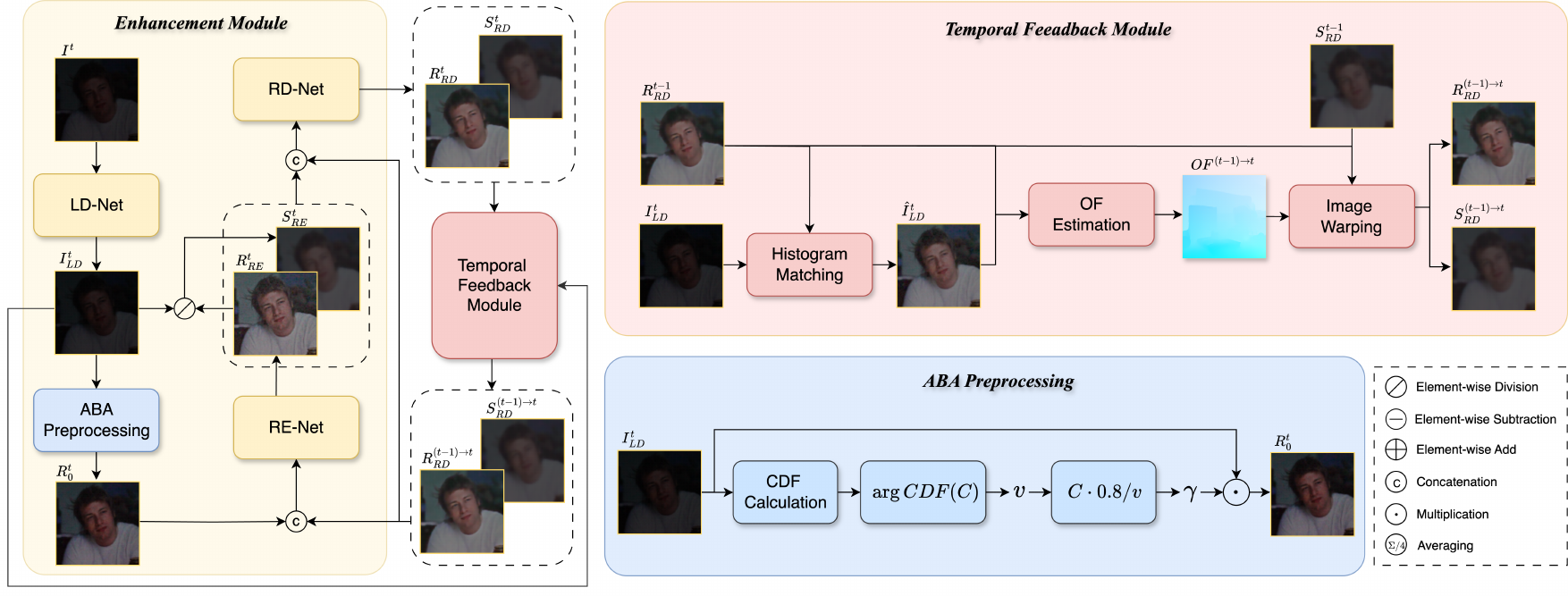}
  \caption{Overall framework of the proposed TempRetinex. The final output is $R^t_{RD}$.}
  \label{fig:framework}
\end{figure*}

\begin{itemize}
    \item We introduce Brightness Consistency Preprocessing (BCP) based on image statistics, enabling robust handling of varying lighting conditions that improves the generalizability of illumination estimation.
    
    \item We propose a multiscale temporal consistency-aware loss function that ensures similarity between frames. An occlusion-aware mask is integrated to handle motion artifacts in complex scenes.
    
    \item To enhance robustness across diverse textures without increasing model complexity, we incorporate a lightweight Self-Ensemble (SE) strategy. 
    
    \item We propose the Reverse Inference (RI) strategy to refine unconverged frames and contribute more temporal information.
    
\end{itemize}


\section{Related work}
\subsection{Low-light Image Enhancement}
Zero-DCE \cite{Zero-DCE} pioneered a zero-reference low-light enhancement learning framework by formulating image-specific curve estimation. EnlightenGAN \cite{jiang2021enlightengan} introduced adversarial learning with attention-guided global–local discriminators. Retinex-based approaches, such as RUAS \cite{liu2021ruas}, Retinexformer \cite{retinexformer} and Zero-IG \cite{Shi:zero:2024}, enhance decomposition robustness through neural representations and illumination-guided denoising. Recent trends emphasize representation learning and multi-modal cues: NeRCo \cite{NerCO_2023_ICCV} explored implicit neural representations and multimodal supervision; CoLIE \cite{Colie2024} achieves zero-shot exposure correction using coordinate-based mapping; and Wakeup-Darkness \cite{zhang2025Wakeup-Darkness} integrates semantic and depth priors for controllable enhancement, further improving detail recovery. While effective for single images, these methods work poorly on video content due to their lack of temporal constraints, which can lead to obvious inter-frame flickering.

\subsection{Low-light Video Enhancement}
Supervised methods are commonly used for low-light video enhancement: SDSDNet \cite{wang2021sdsd} employed a Retinex-based pipeline, BVI-CDM \cite{Lin:Low:2024} leveraged wavelet-based conditional diffusion model for high-fidelity generation, and BVI-Mamba \cite{huang:bvi:2025} integrated the visual state space model to achieve the computational efficiency. Unsupervised methods such as SGZSL \cite{Zheng:semantic:2022}, incorporate semantic guidance, while Zero-TIG \cite{li2025zerotig}, introduces recursive temporal feedback. However, all of these methods lack comprehensive temporal modeling and hence fail in the presence of complex motions or illuminations.


\section{Methodology}

As illustrated in \cref{fig:framework}, TempRetinex includes LD-Net, RE-Net and RD-Net as the enhancement module. BCP is introduced in \cref{subsec:BCP} to adjust histogram distributions. The temporal feedback module employs optical flow (OF) estimation for inter-frame alignment. The final output is $R^t_{RD}$.

Our model is grounded in Retinex theory, where a low-light image $I$ is the element-wise product of reflectance $R$ and illumination $S$. According to \cite{zhang2021rethinking}, camera sensors simultaneously capture both signal-independent and signal-dependent noise. The former can be represented as an additive noise $n$, while the latter mainly consists of shot noise $N$ which is related to the incident photon numbers. Following \cite{Shi:zero:2024}, the $S$ acts as a global gain unaffected by noise, whereas $R$, corresponding to reflected photon numbers, is contaminated by noise during the imaging process. To account for real-world conditions, we extend the model as \cref{eq:retinex}.
\begin{align}
 I = (R+N) \circ S + n.
 \label{eq:retinex}
\end{align}

\input{img/histogram}

\subsection{Brightness Consistency Preprocessing}
\label{subsec:BCP}

\noindent\textbf{Generalization problem due to diverse brightness.} As shown in \cref{fig:histogram} (a) and (b), histograms of the same scene captured at 10\% and 20\% brightness (compared to a reference daylight illumination)  vary markedly. While supervised methods can achieve good generalization, they depend on large labeled datasets covering a continuous spectrum range which is impractical. Most unsupervised methods perform poorly under varying brightness levels; results are typically underexposed when inputs are darker and overexposed when inputs are brighter than the training data mean. This is attributed to domain shifts across illumination levels and the absence of ground-truth supervision to correct deviations \cite{zhang2022towards,narayanan2022challenges}.

Global contrast stretching is a commonly used preprocessing method in professional low-light image processing. Inspired by this, we propose BCP, a simple yet effective method based on image statistics and Retinex theory. Since reflectance represents the inherent properties of objects, its distribution should remain stable across brightness levels. Thus, we prioritize an initial reflectance $R_0$ estimation, rather than directly decomposing illumination in \cite{li2025zerotig}. 

BCP first calculates the grayscale value $v$ at which the image cumulative distribution function (CDF) reaches a threshold $C$ (empirically set to 0.98 to exclude extreme outliers such as saturated pixels or noise). Here $v$ represents the maximum valid brightness under current illumination. The histogram amplification coefficient $\gamma$ is then defined as
\begin{align}
\gamma = C \cdot 0.8 / v,  \qquad
R_0 = \gamma \cdot I. 
\label{eq:BCP}
\end{align}

The factor 0.8 here serves as a safety margin: directly scaling to the maximum intensiy may lead to amplification of noise and saturated regions. Moreover, leaving some dynamic range allows the subsequent RE-Net to refine the reflectance adaptively, improving detail recovery. This process explicitly aligns intensity under different exposures to similar distributions as in \cref{fig:histogram} (c) and (d). Defeined in \cref{eq:BCP}, $R_0$ is regarded as the initial reflectance and refined in RE-Net.

\subsection{Network Structure}

\begin{figure}
  \centering
  \includegraphics[width=0.9\linewidth]{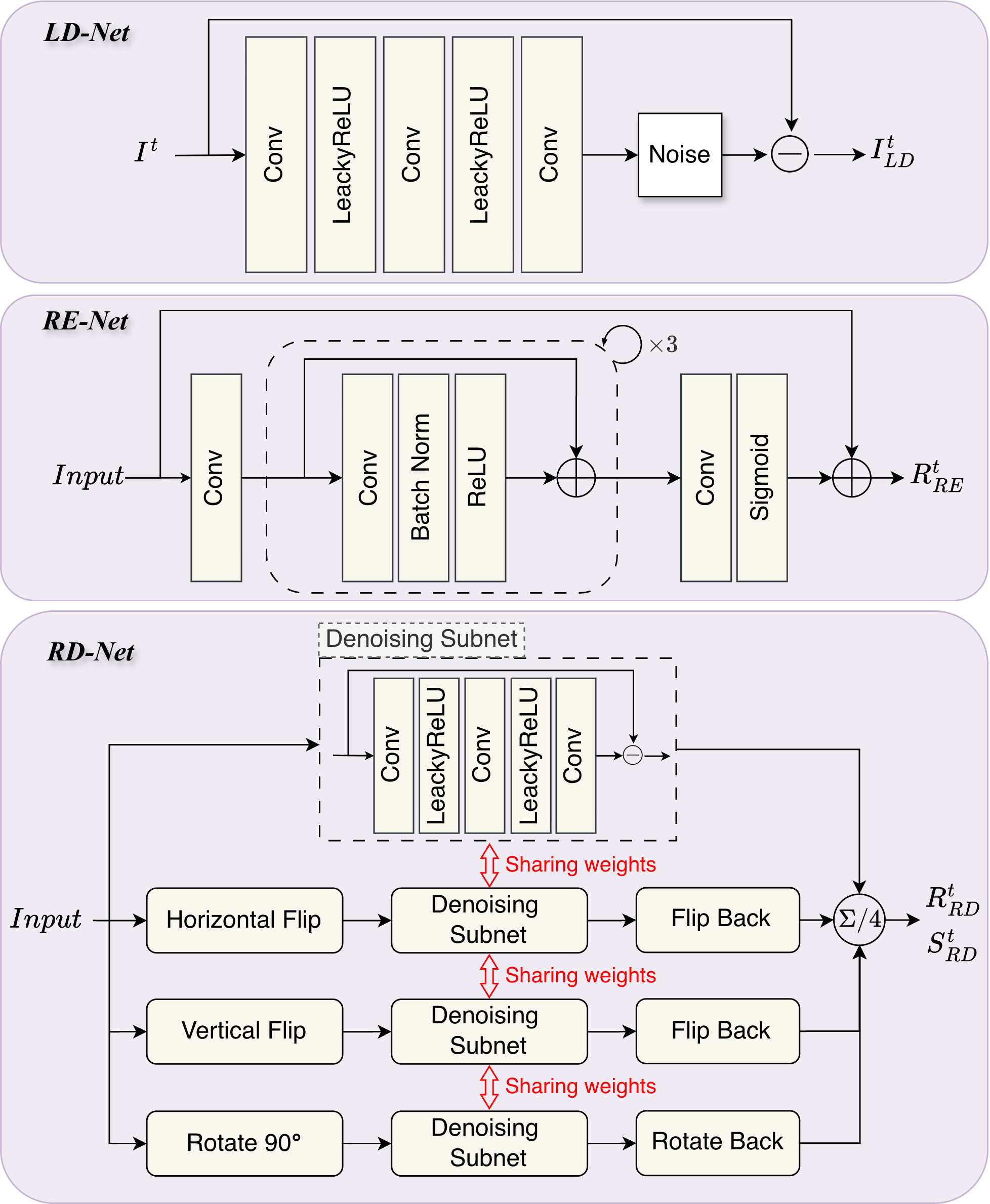}
  \caption{Architectures of LD-Net, RE-Net and RD-Net.}
  \label{fig:subnetworks}
\end{figure}

\subsubsection {Enhancement Module}
\label{sssec:enhance}
As illustrated in \cref{fig:framework}, the Enhancement Module module consists of four core components: a low-light denoising network (LD-Net), the BCP, the reflectance estimation network (RE-Net), and the reflectance denoising network (RD-Net). The subnetwork architectures are shown in \cref{fig:subnetworks}.

Input $I$ is first denoised by LD-Net to generate $I_{LD}$. We apply BCP on $I_{LD}$ to estimate an initial reflectance $R_0$ with a relatively normalized distribution. RE-Net optimizes $R_0$ into $R_{RE}$ using a convolutional network with a residual structure. The illumination $S_{RE}$ is decomposed through element-wise division of $I_{LD}$ by $R_{RE}$. Notably, the $R_{RE}$ corresponds to an ideal reflectance with shot noise $N$, whereas the $S_{RE}$ can be regarded as noise-free due to the smoothness constraint imposed on $S_{RE}$.

$R_{RE}$ and $S_{RE}$ are next concatenated and fed into RD-Net where the SE strategy is introduced. Specifically, it processes four geometric transformations of the input (original, horizontal flip, vertical flip, 90° rotation) through a weight-shared network, and the final denoised output $R_{RD}$ is the average of their inverse-transformed results.

\subsubsection {Temporal Feedback Module}
\label{sssec:tempfeedback}

Video restoration is an ill-posed problem, often leading to inter-frame flickering when processing frames independently without temporal constraints. To this end, we propose an improved temporal feedback module inspired by \cite{li2025zerotig}, achieving inter-frame information fusion through OF alignment technology.

To describe the temporal feedback mechanism, we employ $t$ to denote timestep. This module first estimates motion between adjacent frames. Histogram matching (HM) $\mathcal{M}$ is performed on $I^t_{LD}$ to align its intensity distribution with $R^{t-1}_{RD}$ following \cref{eq:HM}. The processed image $\hat{I}^t_{LD}$ and $R^{t-1}_{RD}$ are used for OF estimation $\mathcal{Q}^{(t-1) \rightarrow t}$ to compute the optical displacement map $OF^{(t-1) \rightarrow t}$ as shown in \cref{eq:RAFT}.
%
\begin{align}
\hat{I}^t_{LD} &= \mathcal{M}(I^t_{LD} \, | \, R^{t-1}_{RD}) \label{eq:HM} \\
OF^{(t-1) \rightarrow t} &= \mathcal{Q}^{(t-1) \rightarrow t}(R^{t-1}_{RD}, \; \hat{I}^t_{LD}) \label{eq:RAFT}
\end{align}

Based on $OF^{(t-1) \rightarrow t}$, the module performs image warping on $R^{t-1}_{RD}$ and $S^{t-1}_{RD}$, obtaining aligned results $R^{(t-1)\rightarrow t}_{RD}$ and $S^{(t-1)\rightarrow t}_{RD}$. For the first frame in a sequence, historical components are initialized as zero vectors. These temporal features are then channel-wise concatenated with current ones for both RE-Net and RD-Net. For OF estimation, we fine-tuned RAFT \cite{teed2020raftrecurrentallpairsfield} on the Sintel \cite{Butler:ECCV:2012} dataset based on the synthetic low light image method proposed in \cite{lin2025towards} to improve the robustness. More details could be referred to the supplementary material.

We also explored architectures based on self-attention and state-space models. However, these introduced substantial computational and memory overheads, while yielding negligible performance gains. Balancing efficiency and effectiveness, we select a convolutional neural network as the backbone.

\begin{table*}[htbp]
  \centering
  \caption{Quantitative comparison on BVI-RLV and DID datasets. Video- and image-based methods are both included. For unsupervised methods, we report results both without (w/o) and with (w/) Histogram Matching (HM). \textbf{Bold} and \underline{underline} denote the best and second-best performances among all \textbf{unsupervised} methods, respectively. $^*$ denotes supervised methods, the supervised results are listed as reference.}
  \label{tab:evaluation}
  
  \setlength{\tabcolsep}{2.5pt}
  \scriptsize
  
  \resizebox{\textwidth}{!}{%
  \begin{tabular}{c|c@{\hskip 7pt}|ccccccc@{\hskip 7pt}|ccccccc}  
    \toprule
    \multicolumn{1}{c}{\multirow{3}{*}{Method}} & \multicolumn{1}{c}{\multirow{3}{*}{Type}} & \multicolumn{7}{c}{BVI-RLV \cite{Lin:BVIRLV:2024}} & \multicolumn{7}{c}{DID \cite{Fu_2023_ICCV}} \\

    \cline{3-16} 
    
    \multicolumn{1}{c}{}& \multicolumn{1}{c}{}& \multicolumn{3}{c}{w/o HM} & \multicolumn{3}{c}{w/ HM} & \multicolumn{1}{c}{\multirow{2}{*}{MABD $\downarrow$}} & \multicolumn{3}{c}{w/o HM} & \multicolumn{3}{c}{w/ HM} & \multirow{2}{*}{MABD $\downarrow$} \\
    
    \cmidrule(lr){3-5} \cmidrule(lr){6-8} \cmidrule(lr){10-12} \cmidrule(lr){13-15} 
    
    \multicolumn{1}{c}{}& \multicolumn{1}{c}{}& PSNR$\uparrow$ & SSIM$\uparrow$ & LPIPS$\downarrow$ & PSNR$\uparrow$ & SSIM$\uparrow$ & LPIPS$\downarrow$ & \multicolumn{1}{c}{} & PSNR$\uparrow$ & SSIM$\uparrow$ & LPIPS$\downarrow$ & PSNR$\uparrow$ & SSIM$\uparrow$ & LPIPS$\downarrow$ & \\
    \midrule
    
    \textcolor{gray}{SDSDNet$^*$ \cite{wang2021sdsd}} & \textcolor{gray}{V} & \textcolor{gray}{20.69} & \textcolor{gray}{0.726} & \textcolor{gray}{0.148} & \textcolor{gray}{--} & \textcolor{gray}{--} & \textcolor{gray}{--} & \textcolor{gray}{--} & \textcolor{gray}{21.88} & \textcolor{gray}{0.834} & \textcolor{gray}{0.216} & \textcolor{gray}{--} & \textcolor{gray}{--} & \textcolor{gray}{--} & \textcolor{gray}{--} \\
    
    \textcolor{gray}{BVI-CDM$^*$ \cite{Lin:Low:2024}} & \textcolor{gray}{V} & \textcolor{gray}{30.51} & \textcolor{gray}{0.888} & \textcolor{gray}{0.089} & \textcolor{gray}{--} & \textcolor{gray}{--} & \textcolor{gray}{--} & \textcolor{gray}{--} & \textcolor{gray}{23.88} & \textcolor{gray}{0.802} & \textcolor{gray}{0.134} & \textcolor{gray}{--} & \textcolor{gray}{--} & \textcolor{gray}{--} & \textcolor{gray}{--} \\
    
    \textcolor{gray}{BVI-Mamba$^*$ \cite{huang:bvi:2025}} & \textcolor{gray}{V} & \textcolor{gray}{31.22} & \textcolor{gray}{0.912} & \textcolor{gray}{0.071} & \textcolor{gray}{--} & \textcolor{gray}{--} & \textcolor{gray}{--} & \textcolor{gray}{--} & \textcolor{gray}{24.21} & \textcolor{gray}{0.850} & \textcolor{gray}{0.168} & \textcolor{gray}{--} & \textcolor{gray}{--} & \textcolor{gray}{--} & \textcolor{gray}{--} \\

    \textcolor{gray}{Retinexformer$^*$ \cite{retinexformer}} & \textcolor{gray}{I} & \textcolor{gray}{32.63} & \textcolor{gray}{0.906} & \textcolor{gray}{0.265} & \textcolor{gray}{--} & \textcolor{gray}{--} & \textcolor{gray}{--} & \textcolor{gray}{--} & \textcolor{gray}{24.15} & \textcolor{gray}{0.849} & \textcolor{gray}{0.216} & \textcolor{gray}{--} & \textcolor{gray}{--} & \textcolor{gray}{--} & \textcolor{gray}{--} \\
    \midrule
    
    Zero-DCE \cite{Zero-DCE} & I & 10.54 & 0.430 & 0.528 & 18.93 & 0.488 & 0.507 & 16.05 & 14.08 & 0.686 & 0.404 & 20.13 & 0.655 & 0.462 & 10.62 \\
    RUAS \cite{liu2021ruas} & I & 15.31 & 0.631 & 0.481 & 18.52 & 0.712 & 0.515 & 6.70 & 17.23 & 0.775 & 0.430 & 22.11 & 0.836 & 0.433 & 10.33 \\
    EnlightenGAN \cite{jiang2021enlightengan} & I & 15.49 & 0.518 & 0.515 & 17.88 & 0.550 & 0.522 & 12.26 & 19.37 & 0.725 & 0.403 & 22.93 & 0.754 & 0.402 & 14.20 \\
    NeRCo \cite{NerCO_2023_ICCV} & I & 21.05 & 0.705 & 0.431 & 27.67 & 0.822 & 0.402 & 7.39 & \underline{20.90} & 0.826 & 0.386 & 28.47 & 0.879 & 0.372 & 10.32 \\
    CoLIE \cite{Colie2024} & I & 18.21 & 0.583 & 0.410 & 24.15 & 0.646 & 0.431 & 11.50 & 18.80 & 0.784 & 0.356 & 26.36 & 0.825 & 0.348 & 12.44 \\
    Zero-IG \cite{Shi:zero:2024} & I & 19.37 & 0.639 & 0.398 & 27.84 & 0.834 & 0.370 & \textbf{5.27} & 14.25 & 0.785 & 0.380 & 25.84 & 0.854 & 0.386 & \underline{7.03} \\
    Wakeup-Darkness \cite{zhang2025Wakeup-Darkness} & I & 19.83 & 0.734 & \underline{0.342} & 28.43 & 0.774 & 0.348 & 8.84 & \textbf{20.93} & 0.801 & 0.323 & 30.12 & 0.852 & 0.324 & 11.11 \\
    SGZSL \cite{Zheng:semantic:2022} & V & 13.42 & 0.577 & 0.420 & 24.03 & 0.723 & 0.380 & 12.17 & 15.55 & 0.733 & 0.363 & 25.05 & 0.836 & 0.346 & 11.68 \\
    Zero-TIG \cite{li2025zerotig} & V & 19.34 & 0.790 & 0.360 & 28.05 & 0.854 & 0.368 & 6.27 & 16.79 & 0.818 & 0.373 & 25.96 & 0.836 & 0.410 & \textbf{6.78} \\
    \midrule
    Ours w/o RI & V & \underline{25.02} & \underline{0.824} & 0.344 & \underline{30.16} & \underline{0.874} & \underline{0.331} & 5.33 & 19.26 & \underline{0.847} & \underline{0.319} & \underline{32.03} & \underline{0.905} & \underline{0.288} & 9.67 \\
    Ours w/ RI & V & \textbf{25.11} & \textbf{0.829} & \textbf{0.338} & \textbf{30.27} & \textbf{0.877} & \textbf{0.327} & \underline{5.29} & 19.27 & \textbf{0.849} & \textbf{0.314} & \textbf{32.24} & \textbf{0.908} & \textbf{0.283} & 9.65 \\
    \bottomrule
  \end{tabular}%
  }
\end{table*}
\begin{table}[htbp]
  \centering
    \caption{Ablation study of the contribution of BCP and SE.} 
  \resizebox{\columnwidth}{!}{
  \begin{tabular}{@{}c|cccccc@{}}
    \toprule

    \multicolumn{1}{c}{\multirow{2}{*}{Method}} & \multicolumn{2}{c}{PSNR $\uparrow$} & \multicolumn{2}{c}{SSIM $\uparrow$} & \multicolumn{2}{c}{LPIPS $\downarrow$} \\
    \cmidrule(lr){2-3} \cmidrule(lr){4-5} \cmidrule(lr){6-7} 
       \multicolumn{1}{c}{} & w/o HM & w/ HM & w/o HM & w/ HM & w/o HM & w/ HM \\
        \midrule
        Ours w/o BCP & 20.069 & 29.345 & 0.778 & \underline{0.866} & 0.381 & 0.360 \\
        Ours w/o SE & \underline{24.699} & \underline{29.808} & \underline{0.806} & 0.863 & \underline{0.359} & \underline{0.345} \\
        Ours & \textbf{25.001} & \textbf{30.118} & \textbf{0.822} & \textbf{0.872} & \textbf{0.345} & \textbf{0.333} \\
    \bottomrule
  \end{tabular}}

  \label{tab:ablation}
\end{table}


\subsection{Loss Functions}
\label{subsec:lossfuc}
In addition to the non-reference loss set used in Zero-TIG \cite{li2025zerotig} (refer to this paper for details), we introduce a novel multiscale temporal consistency–aware loss to form a more comprehensive loss function system.

\noindent \textbf{Multi-scale Temporal Consistency-aware Loss.} This loss $L_{mtc}$ is proposed to further strengthen the cross-frame coherence. It minimizes the L1 loss between current reflectance $R^t_{RD}$ and warped reference $R^{(t-1) \rightarrow t}_{RD}$. Considering that OF estimation can only model 2D pixel displacement but fails in spatial structural changes in depth dimension, we introduce an occlusion mask inspired by \cite{zhu2024unrolled}, which is computed by the exponent of square error between $R^t_{RE}$ and $R^{(t-1) \rightarrow t}_{RD}$:
\begin{align}
M=exp(-\omega(||R^t_{RE} -  R^{(t-1) \rightarrow t}_{RD} ||_2^2)),
\label{eq:OccuMask}
\end{align}
where $\omega$ is empirically set to 100. Following findings that downsampled sequences reduce motion blur \cite{shrivastava2023videodynamicspriorinternal}, we apply a spatial pyramid for multiscale supervision: 
\begin{align}
L_{mtc} = \sum_{i=1,2,3,4} \mathcal{D}_i(M \circ(R^t_{RD} -  R^{(t-1) \rightarrow t}_{RD}) ),
\label{eq:L_mtc}
\end{align}
where $\mathcal{D}_i$ denotes bicubic downsampling operation at level i, $\circ$ denotes element-wise multiplication.

\subsection{Reverse Inference Strategy}
\label{sec:reverse_infernce}

Our temporal feedback module utilizes only past frames. While suitable for online processing, this approach cannot leverage future frames and requires a convergence window. For offline applications, we processes the video sequence both forward and backward during inference, then average the results. This strategy differs from bidirectional structure but provides a practical way, which requires no network retraining and remains compatible with forward-only processing.

\input{img/BVI_S17_clip}

\input{img/DID_V51}

\section{Experiments}
\subsection{Implementation Details}
We evaluated our method on two paired datasets: BVI-RLV \cite{Lin:BVIRLV:2024} provides 40 dynamic scenes recorded at 10\%, 20\%, and 100\% brightness (HD video pairs); DID \cite{Fu_2023_ICCV} offers 413 HD pairs under diverse real-world illumination. We use the Adam optimizer ($\beta_1=0.9$, $\beta_2=0.999$), with weight decay $3\times10^{-4}$ and learning rate $5\times10^{-5}$. RAFT inputs are downsampled by $3\times$ during training for efficiency, while full resolution during inference for finer structural details.

\subsection{Benchmark Evaluation}
We compared with both image- and video-based state-of-the-art (SOTA) approaches, including supervised ones as upper bound. All methods were retrained for fairness, except NeRCo, for which we used the pre-trained one due to out-of-memory issues. Evaluation employs PSNR, SSIM, LPIPS \cite{zhang2018perceptual}, and MABD \cite{Jiang:learn:2019}. As brightness could be subjective, we apply HM to align outputs with ground truth to remove brightness variations, thus better assess denoising performance.

As shown in \cref{tab:evaluation}, TempRetinex outperforms existing unsupervised methods on most metrics and remains competitive in PSNR on DID before HM. It can be observed that RI brings consistent (albeit modest) gains across all metrics. Although MABD is slightly higher than Zero-IG \cite{Shi:zero:2024} and Zero-TIG \cite{li2025zerotig}, this is likely due to their darker outputs reducing the absolute brightness changes and hence lowering MABD. In contrast, TempRetinex enhances image brightness while maintaining competitive temporal stability. Qualitative results (\cref{fig:BVI_S17_clip,fig:DID_V51}) demonstrate the superior detail preservation, denoising and temporal stability of our method. 

\begin{figure}
    \centering
    \includegraphics[width=\linewidth]{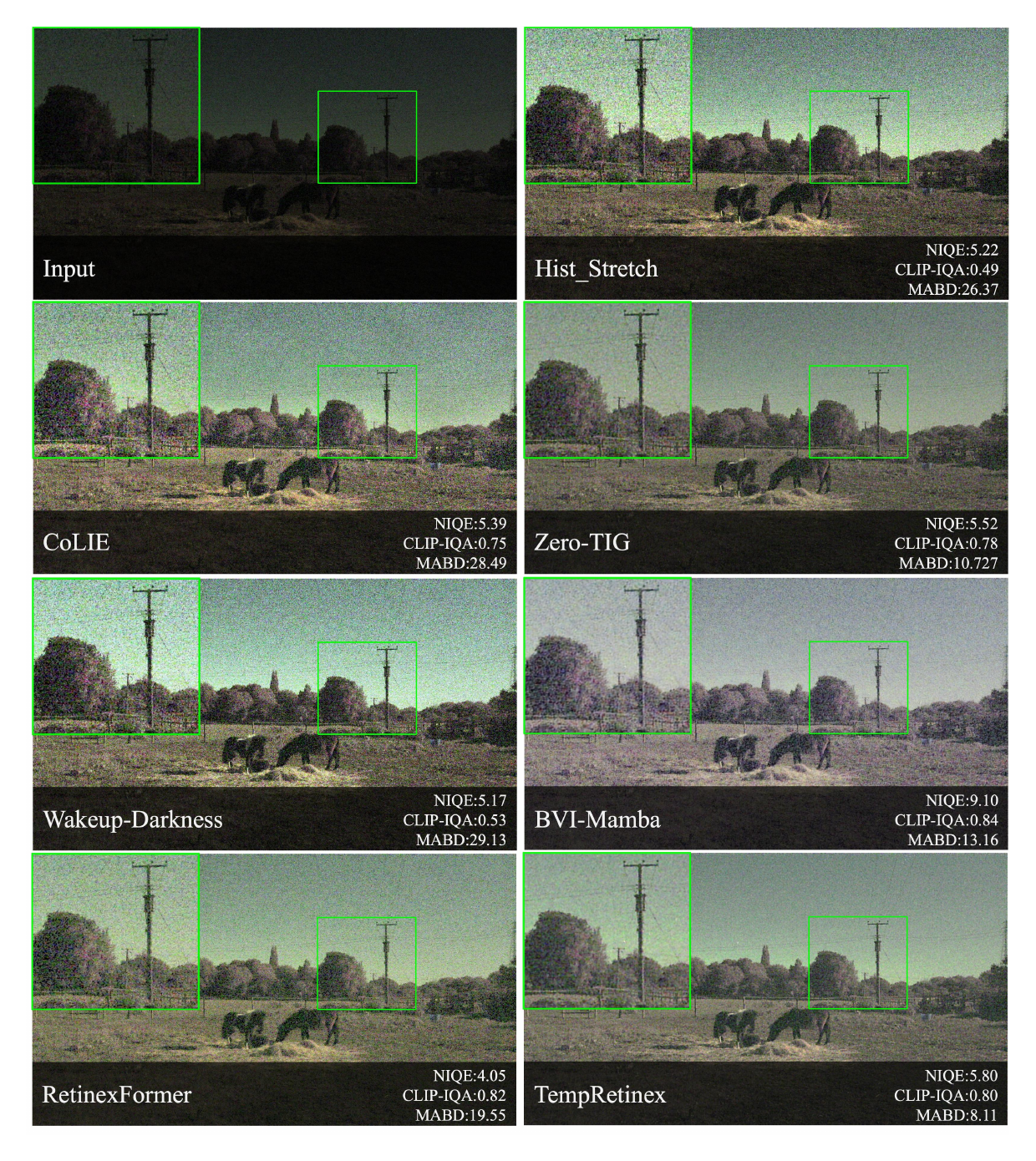}
    \caption{Real moonlit scene enhancement comparison.}
    \label{fig:horse_cmp}
\end{figure}


\subsection{Outdoor low-light video performance}
To verify robustness in real scenes, we test our approach on a moonlit video captured using a Canon ML-105. No-reference image quality metrics (NIQE \cite{NIQE} and CLIP-IQA \cite{wang2022exploring}) are used. We compare against best-performing unsupervised and supervised methods on the BVI-RLV, using pre-trained models for inference. Furthermore, we add BSP results (denoted as \textit{Hist\_Stretch}) for denoising comparison. As shown in \cref{fig:horse_cmp}, TempRetinex delivers stronger noise suppression, and temporal stability. However, we note that our method performs slightly worse in terms of image contrast restoration, which indicates the need for optimization in future work. 

\begin{table}[htbp]
\centering
\caption{MABD on a video with and without $L_{mtc}$.}
\begin{tabular}{c|c}
\toprule
Method & MABD $\downarrow$\\
\midrule
w/o $L_{mtc}$ & 3.2 \\
w/ $L_{mtc}$ & 2.7 \\
\bottomrule
\end{tabular}
\label{tab:MABD}
\end{table}

\begin{figure}[t]
\centering
\includegraphics[width=0.9\linewidth]{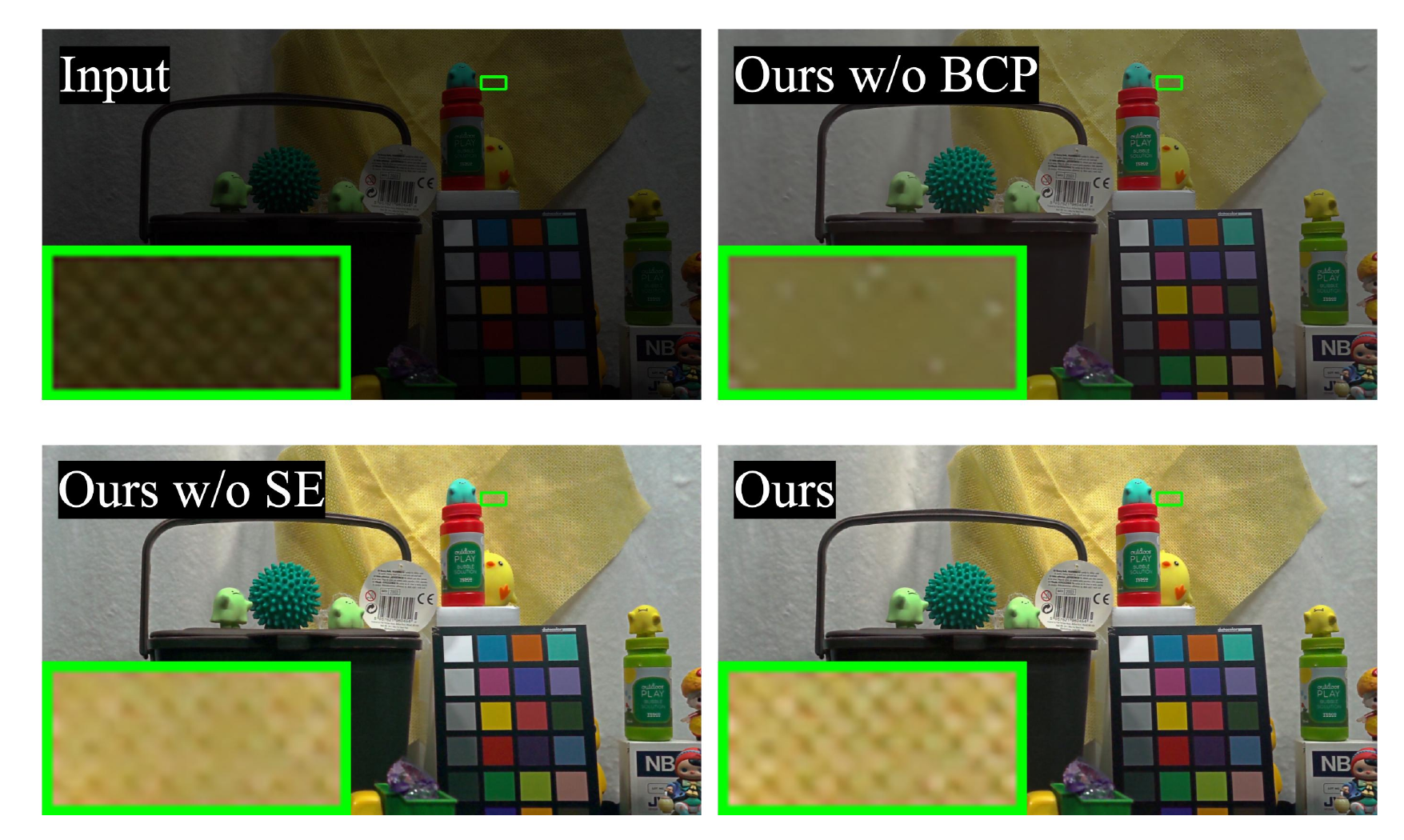}
\caption{Visual comparison of ablation study of BCP and SE.}
\label{fig:ablation_struct}
\end{figure}


\subsection{Ablation Study}
\noindent\textbf{BCP and SE.} \cref{tab:ablation} and \cref{fig:ablation_struct} present the contributions of different network structures on BVI-RLV. As the RI strategy only reorders input frames without changing the network, our ablation studies are reported without it, which does not affect the evaluation of other components. The SE removal clearly degrades denoising performance, while BCP removal causes notable deterioration in metrics, which confirms the importance of distribution standardization.

\noindent\textbf{Multi-scale Temporal Consistency-aware Loss.}  We use the averaged MABD to evaluate $L_{mtc}$ on a representative video sequence \textit{S03\_animals2} in BVI-RLV. \cref{tab:MABD} shows the effectiveness of $L_{mtc}$ in improving temporal consistency. Further analysis is provided in the supplementary material.

\section{Conclusion}

We present TempRetinex, an unsupervised framework for low-light video enhancement. Key contributions include: i) a BCP module for generalizability across diverse brightness; ii) a multiscale temporal consistency-aware loss function with occlusion-aware masks; and iii) the SE and RI strategies are introduced for denoising. Both qualitative and quantitative experiments demonstrate SOTA performance, achieving over 5\% improvement in SSIM, without the need for paired data.

\bibliographystyle{IEEEbib}
\bibliography{icme2025references.bib}

\clearpage

\twocolumn[
\centering
{\LARGE \textbf{TempRetinex: Supplementary Material}\par}
\vspace{0.5em}
]

\renewcommand{\thesection}{S\arabic{section}}

\begin{abstract}
This supplementary material provides extended experiments and implementation details for our unsupervised low-light video enhancement framework, TempRetinex. Additional visual results and quantitative evaluations further validate the effectiveness and robustness of the proposed approach.
\end{abstract}

\section{Generalization problem and more results}
\label{sec:res}

\begin{figure*}
  \centering
  \includegraphics[width=\linewidth]{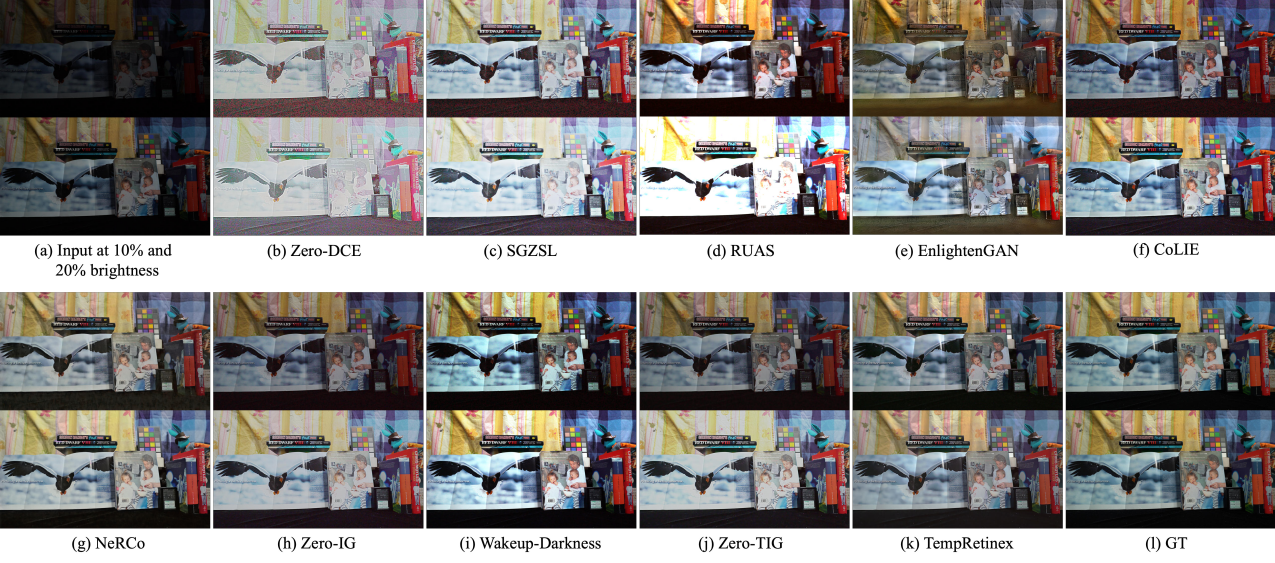}
  \caption{Visual comparison of video \textit{S20\_books3} in BVI-RLV.}
  \label{fig:bvi_s20_cmp}
\end{figure*}

\begin{figure*}
  \centering
  \includegraphics[width=\linewidth]{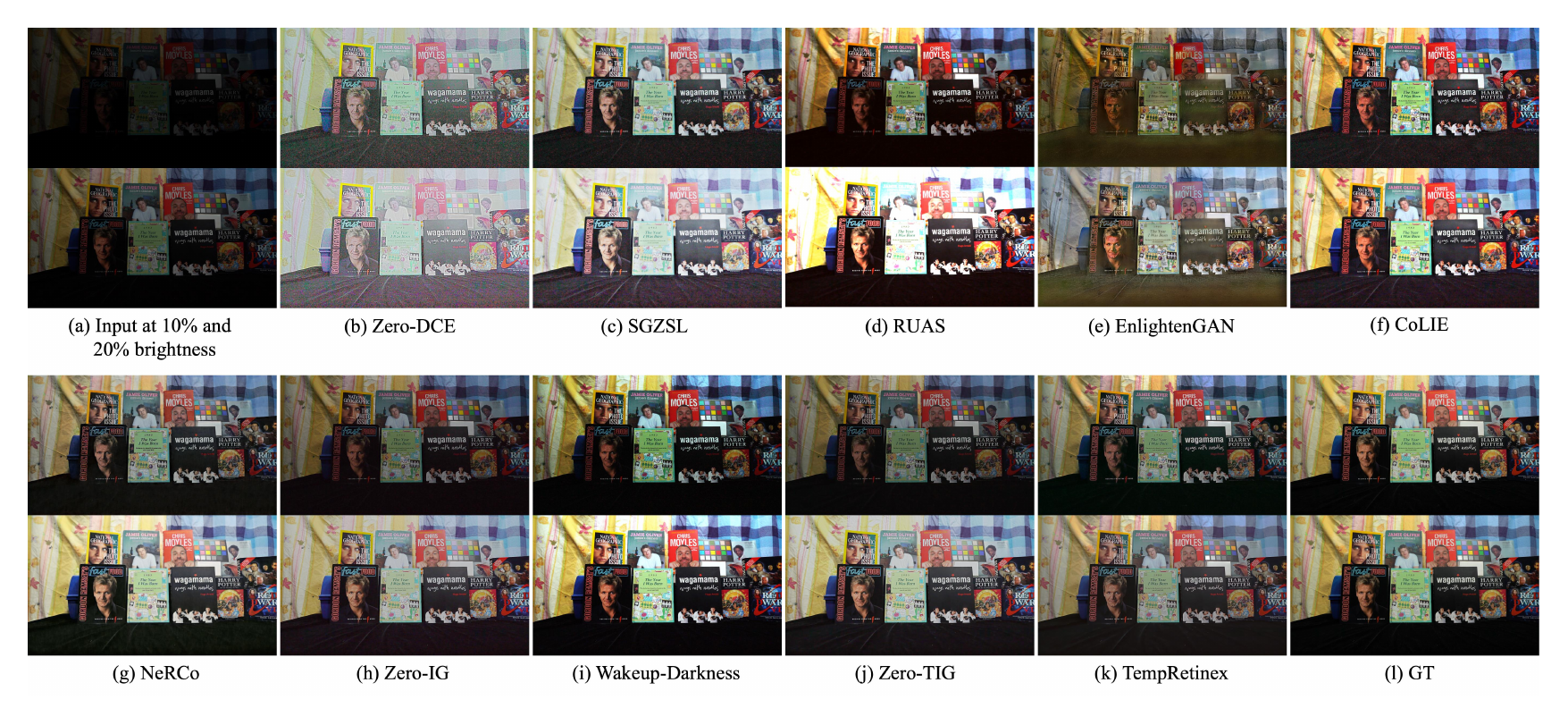}
  \caption{Visual comparison of video \textit{S17\_faces2} in BVI-RLV.}
  \label{fig:bvi_s17_cmp}
\end{figure*}

In the main paper, we have discussed the issue that existing unsupervised methods produce inconsistent results under varying lighting conditions. Due to space constraints, only cropped visual examples were included. In this section, we provide more comprehensive results to further illustrate this generalization problem and compare them with the outputs of our method, as shown in \cref{fig:bvi_s20_cmp} and \cref{fig:bvi_s17_cmp}. All inputs are drawn from the 10\% and 20\% brightness sequences of the BVI-RLV dataset.


SGZSL and RUAS exhibit overexposure, while EnlightenGAN and NeRCo produce artifacts. Although Zero-IG and Zero-TIG improve brightness, the text in the images is blurry which show poorer contrast. CoLIE and Wakeup-Darkness suffer from noticeable noise. By contrast, TempRetinex preserves details with better luminance and color fidelity. In terms of brightness consistency, other methods' outputs show visible inconsistency. Notably, Zero-IG and Zero-TIG produce underexposed images at 10\% brightness.

\section{Design of RE-Net}
\label{sec:re-net}

The structure of RE-Net can be referred to the main paper. Inspired by the stage-wise optimization strategy in \cite{liu2021ruas,ma2022toward}, RE-Net adopts a convolutional network with residual structure for iterative refinement, following the update equations:
\begin{align}
\mathcal{F}(R_k): \begin{cases}
    u_k= \mathcal{G}(R_k) , \\
    R_{k+1} = R_k+u_k ,
\end{cases}
\label{eq:Riter}
\end{align}
where $\mathcal{G}$ denotes the convolutional mapping layers, $k$ represents the iteration index, and $u_k$ indicates the residual term at stage $k$. $R_0$ is the initial estimate for the RE-Net's optimization process. The Retinex decomposition is computed as \cref{eq:RE}:
\begin{align} 
\begin{split}
& R_{RE} = \mathcal{F}_{RE}(R_0), \\
& S_{RE} = I_{LD} \; \oslash \; R_{RE},
\end{split}
\label{eq:RE}
\end{align}
where $\oslash$ denotes element-wise division and $\mathcal{F}_{RE}$ is the estimation function. Notably, the $R_{RE}$ here corresponds to an ideal reflectance with sensor noise $N$, whereas the $S_{RE}$ can be regarded as noise-free due to the smoothness constraint imposed on $S_{RE}$ as described in \cref{eq:L_smooth}.

In our implementation, we set $k=1$ empirically. An ablation study was conducted by increasing $k$ to 2 and 3. \cref{tab:re-net} demonstrates that larger $k$ values do not improve performance but correspondingly increase computational cost (FLOPs are reported for HD image size). The model contains 0.13M parameters.

\begin{table}
  \centering
  \caption{Ablation study on the number of refinement stages ($k$) in RE-Net. All results are reported without the RI strategy and histogram matching.}
  \label{tab:re-net} 
  \resizebox{\columnwidth}{!}{
    \begin{tabular}{@{}lcccc@{}}
      \toprule
       & PSNR $\uparrow$ & SSIM $\uparrow$ & LPIPS $\downarrow$ & FLOPs ($\times 10^{12}$) $\downarrow$ \\
      \midrule
      $k=1$ (Ours) & 25.00 & 0.822 & 0.345 & 2.40 \\
      $k=2$ & 24.85 & 0.811 & 0.355 & 4.28 \\
      $k=3$ & 25.29 & 0.809 & 0.365 & 6.15 \\
      \bottomrule
    \end{tabular}
  }
\end{table}

\section{Fine-tuning of RAFT}
\label{sec:raft}

\begin{figure}
  \centering
  \includegraphics[width=\linewidth]{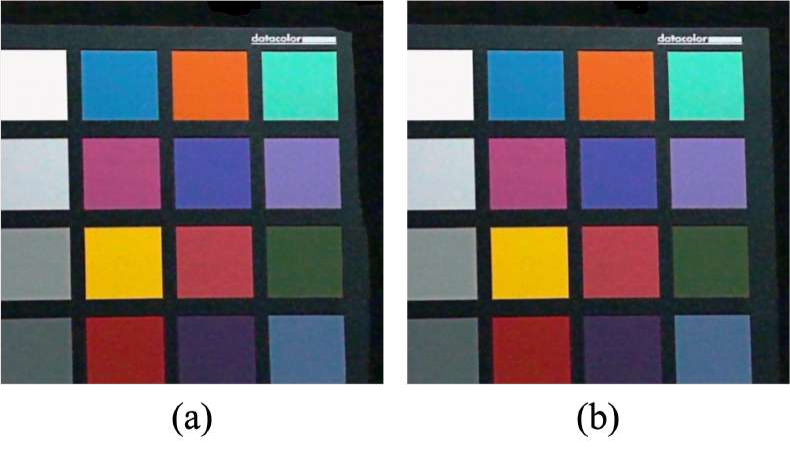}
  \caption{(a) Warping result using the original pre-trained RAFT. (b) Warping result using our fine-tuned RAFT.}
  \label{fig:of_finetune}
\end{figure}

The original RAFT model was pre-trained on clean, high-contrast data. In our application, however, the current frame is a noisy low-light image. Although we perform histogram matching with the previous enhanced frame, contrast inconsistency between frames may still persist. Therefore, we fine-tune RAFT on the Sintel dataset \cite{Butler:ECCV:2012} by applying random noise and brightness degradation as described in \cite{lin2025towards}. As shown in \cref{tab:of_finetune}, this fine-tuning leads to significant improvements across all metrics. Furthermore, \cref{fig:of_finetune} compares the warping results of the original and fine-tuned models on the same frame, demonstrating that the adapted model yields more reliable optical flow under low-light conditions.

\begin{table}
  \centering
  \caption{Comparison of optical flow estimation using the original pre-trained RAFT and our fine-tuned version on low-light data. Results are reported with and without histogram matching (HM).}
  \label{tab:of_finetune}
  \resizebox{\columnwidth}{!}{
    \begin{tabular}{@{}lcccccc@{}}
      \toprule
      & \multicolumn{2}{c}{PSNR $\uparrow$} & \multicolumn{2}{c}{SSIM $\uparrow$} & \multicolumn{2}{c}{LPIPS $\downarrow$} \\
      \cmidrule(lr){2-3} \cmidrule(lr){4-5} \cmidrule(lr){6-7}
      Method & w/o HM & w/ HM & w/o HM & w/ HM & w/o HM & w/ HM \\
      \midrule
      Pre-trained & 25.00 & 30.112 & 0.822 & 0.872 & 0.345 & 0.333 \\
      Fine-tuned (Ours)  & 25.02 & 30.162 & 0.824 & 0.874 & 0.344 & 0.331 \\
      \bottomrule
    \end{tabular}
  }
\end{table}

\section{Loss Functions}
\label{sec:loss}

\subsection{Loss terms}
For LD-Net, we adopt the same optimization strategy as \cite{mansour2023zero} by decomposing the noisy image into two subimages through downsamplers $G_1$ and $G_2$. Given $\mathcal{F}_{LD}$ as the noise prediction function, and $I$ as $I_{LD}$ in short, the residual loss $L_{res1}$ and consistency loss $L_{cons1}$ are defined as \cref{eq:L_res1} and \cref{eq:L_cons1}.
\begin{align}
L_{res1} &= ||G_1(I) - \mathcal{F}_{LD}(G_1(I)) - G_2(I)||^2_2 \nonumber \\
         &+ ||G_2(I) - \mathcal{F}_{LD}(G_2(I)) - G_1(I)||^2_2\, .
\label{eq:L_res1}
\end{align}
\vspace{-8mm}
\begin{align}
L_{cons1}&=||G_1(I) - \mathcal{F}_{LD}(G_1(I)) - G_1(I - \mathcal{F}_{LD}(I))||^2_2 \nonumber \\
          &+ ||G_2(I) - \mathcal{F}_{LD}(G_2(I)) - G_2(I - \mathcal{F}_{LD}(I))||^2_2\ .  
\label{eq:L_cons1}
\end{align}

For RE-Net, we introduce three optimization objectives to refine both reflectance and illumination. Concretely, the global brightness constraint $L_{glob}$ regulates the mean intensity of $R^t_{RE}$ to approach a predefined upscaling factor $\alpha$:
\begin{align}
L_{glob} = ||R^t_{RE} - \alpha I^t_{LD}||^2_2,
\label{eq:L_glob}
\end{align}
where $\alpha=Y_H\cdot Y_L^{-1}$, $Y_L$ represents the mean value of the luminance of $I^t_{LD}$, while $Y_H$ is the mean value of the luminance of normal brightness images.

The pixel-wise adjustment loss $L_{pix}$ establishes nonlinear mapping relationships between different intensity levels to achieve pixel-wise brightness adjustment. $L_{pix}$ is depicted in \cref{eq:L_pix}, where the scaling factor $\lambda$ is set to $\alpha^{-1}0.7^{-\alpha}$ according to \cite{li2025zerotig}.
\begin{align}
L_{pix}=||S^t_{RE} - \lambda(\alpha I^t_{LD})^{\alpha} ||^2_2.
\label{eq:L_pix}
\end{align}

Based on the prior assumption that illumination should be continuous, $L_{s}$ applies TV regularization to ensure spatial continuity of the illumination:
\begin{align}
L_{s} = (|\nabla_x S^t_{RE}| + |\nabla_y S^t_{RE}|)^2 + \sum_{i,j} w_{ij} |S^t_{RE,i} - S^t_{RE,j}|,
\label{eq:L_smooth}
\end{align}
where \(\nabla_x\) and \(\nabla_y\) represent horizontal and vertical gradient operators, and \(w_{i,j}\) are Gaussian-weighted coefficients for a 5×5 neighborhood \(\mathcal{N}(i)\) around pixel \(i\). 

For RD-Net, we employ the same downsampling strategy along with loss functions $L_{res2}$ and consistency loss $L_{cons2}$ to train the denoising performance for $R^t_{RD}$. The illumination consistency loss $L_{ill}$ as in \cref{eq:L_ill}, minimizes the mean square error between $S^t_{RD}$ and $S^t_{RE}$ to maintain stability before and after denoising. 
\begin{align}
L_{ill}= \| S^t_{RD} - S^t_{RE} \|_2^2 .
\label{eq:L_ill}
\end{align}

Following \cite{Shi:zero:2024}, we incorporate $L_{var}$ and $L_{color}$ to preserve texture and color fidelity in $R^t_{RD}$, while their interactive denoising loss $L_{inter}$ further enhances performance. Implementation details of these components are available in the original work.

The multi-scale temporal consistency-aware loss $L_{mtc}$ is described in detail in the main paper.

In summary, the total loss function is defined as \(L_{total}=L_{res1}+L_{cons1}+L_{glob}+L_{pix}+L_{s}+L_{res2}+L_{cons2}+L_{ill}+L_{inter}+L_{var}+L_{color}+L_{mtc}\).

\subsection{Impact of $L_{mtc}$}
\label{ssec:Lmtc}
\begin{figure}[t]
\centering
\includegraphics[width=0.8\linewidth]{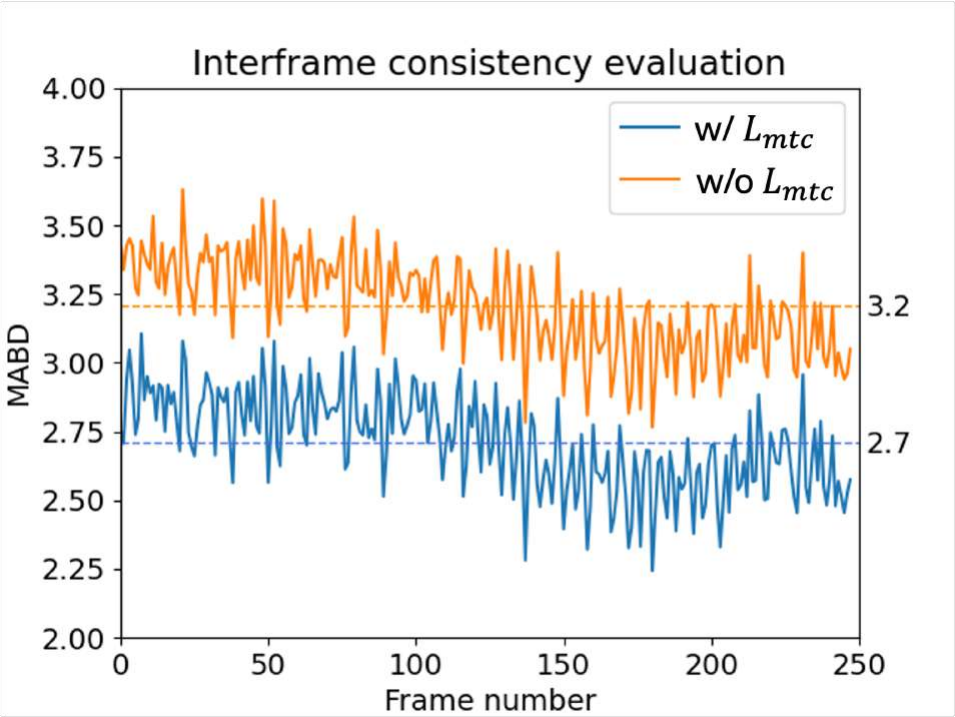}
\caption{MABD on \textit{S03\_animals2} with and without $L_{mtc}$.}
\label{fig:MABD}
\end{figure}

While the effect of $L_{mtc}$ has been validated in the ablation study in the main paper, we further illustrate its impact by reporting inter-frame MABD on the sequence \textit{S03\_animals2} in \cref{fig:MABD}. The results show that $L_{mtc}$ effectively reduces inter-frame variations across the entire video.

\subsection{Impact of $\omega$}
\label{ssec:omega}
\begin{figure}
  \centering
  \includegraphics[width=0.9\linewidth]{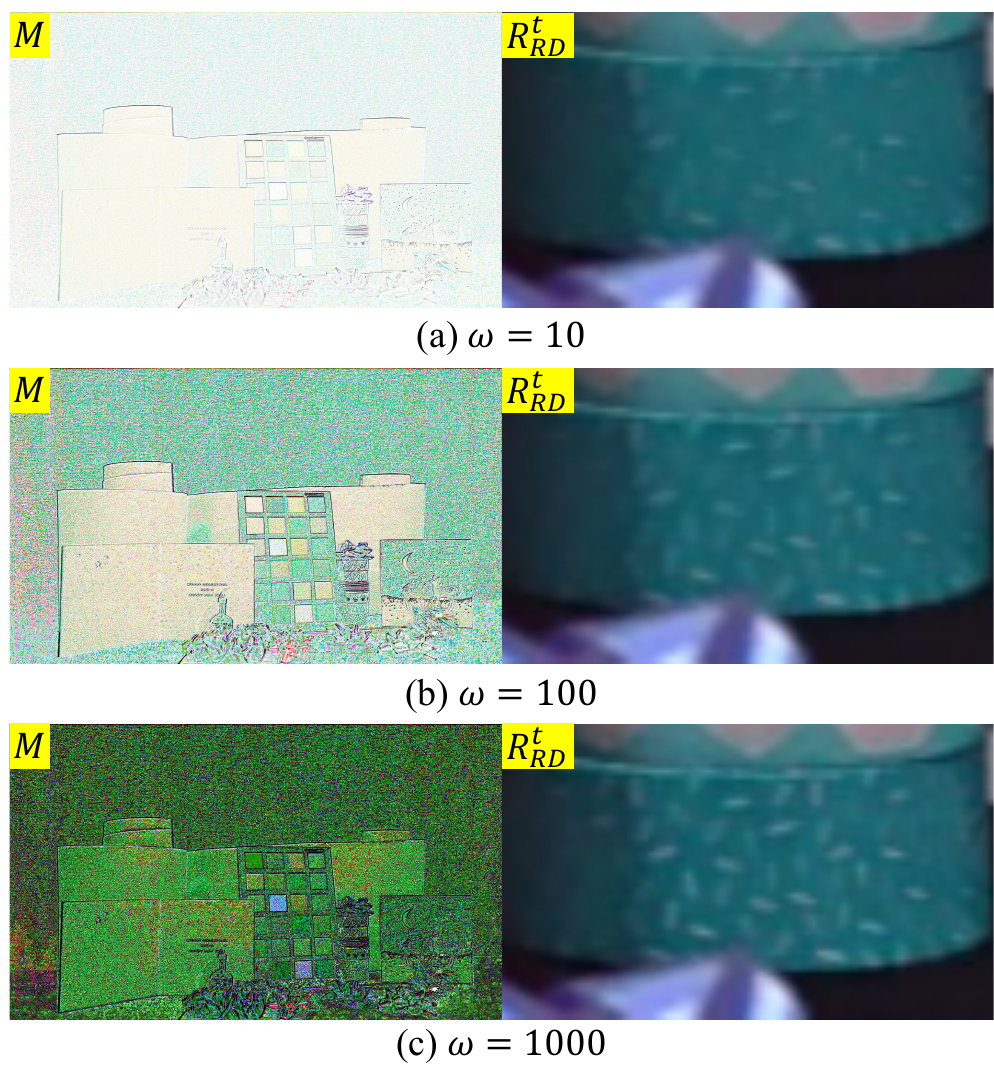}
  \caption{Visualization of occlusion masks and outputs under different $\omega$}
  \label{fig:omega}
\end{figure}

In the temporal loss, we employ an occlusion mask $M$ to mitigate errors in optical flow estimation, where a hyperparameter $\omega$ controls the mask intensity. As illustrated in \cref{fig:omega}, we visualize the pixel-wise magnitude of $M$ under $\omega = 10, 100, 1000$. A larger mask value indicates stronger reliance on the previous frame. A small $\omega$ yields higher dependence on historical frames but risks propagating flow-warping errors; conversely, a large $\omega$ reduces such reliance but also diminishes temporal smoothing. \cref{tab:omega} presents the impact of hyperparameter $\omega$, demonstrating a clear trade-off between enhancement quality and temporal stability. We therefore set $\omega = 100$ as a balanced trade-off between accuracy and temporal coherence.

\begin{table}
  \centering
  \caption{Ablation study on the hyperparameter $\omega$. All results are reported without the RI strategy and histogram matching.}
  \label{tab:omega} 
  \resizebox{\columnwidth}{!}{
    \begin{tabular}{@{}lcccc@{}}
      \toprule
       & PSNR $\uparrow$ & SSIM $\uparrow$ & LPIPS $\downarrow$ & MABD $\downarrow$ \\
      \midrule
      $\omega=10$ & 25.06 & 0.825 & 0.342 & 5.14 \\
      $\omega=100$ (Ours) & 25.00 & 0.822 & 0.345 & 5.35 \\
      $\omega=1000$ & 25.35 & 0.807 & 0.323 & 5.75 \\
      \bottomrule
    \end{tabular}
  }
\end{table}

\subsection{Contribution of each term}

Our loss function comprises 12 terms. To evaluate the contribution of each, we performed an ablation study by removing them one at a time. The radar chart below illustrates the changes in PSNR and SSIM on the BVI-RLV dataset after the removal of each loss term.

The ablation results show that every loss term contributed positively to the overall performance. The global brightness constraint \(L_{\text{glob}}\) was the most critical; its removal caused PSNR to plummet to 5.001, confirming its fundamental role in maintaining a proper luminance distribution. Removing the color loss \(L_{\text{color}}\) and the interactive denoising loss \(L_{\text{inter}}\) significantly reduced SSIM, underscoring their importance for preserving color naturalness and structural consistency.

\begin{figure}
  \centering
  \includegraphics[width=0.95\linewidth]{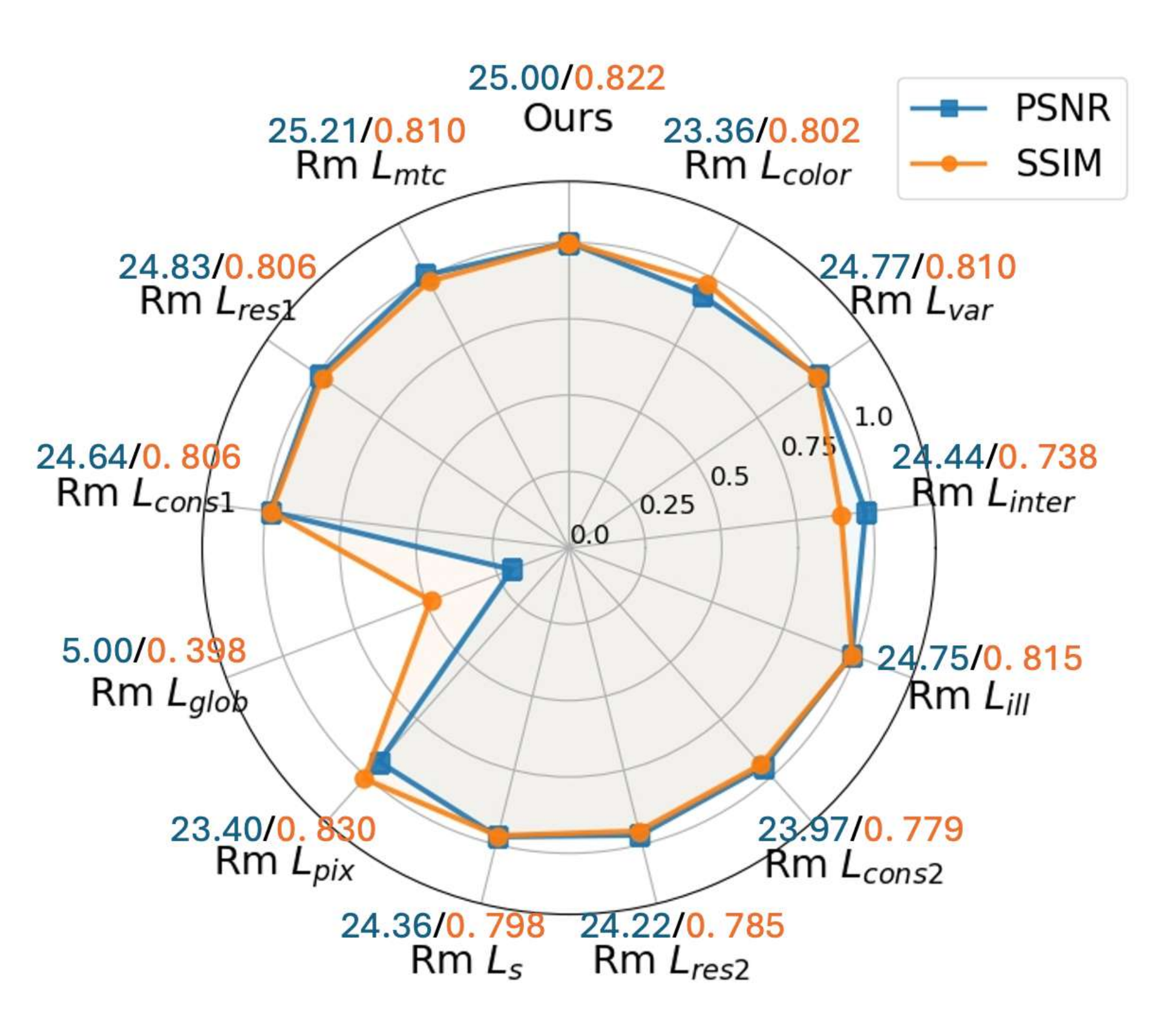}
  \caption{PSNR and SSIM radar charts with values normalized to our complete model (Ours = 1.0).}
  \label{fig:radar}
\end{figure}

Notably, removing the pixel-wise luminance adjustment loss \(L_{\text{pix}}\) increased SSIM slightly while lowering PSNR, suggesting that this term may enhance structural naturalness at a minor cost to pixel-level fidelity. Conversely, although removing the multi-scale temporal consistency loss \(L_{\text{mtc}}\) yielded a slight PSNR gain, it decreased SSIM and increased temporal noise. As analyzed in \cref{ssec:omega}, this is equivalent to setting the mask intensity \(\omega\) to infinity, which reduces temporal smoothness and amplifies flickering artifacts.

\section{Reverse Inference}
\label{sec:RI}
\begin{figure}
    \centering
    \includegraphics[width=\linewidth]{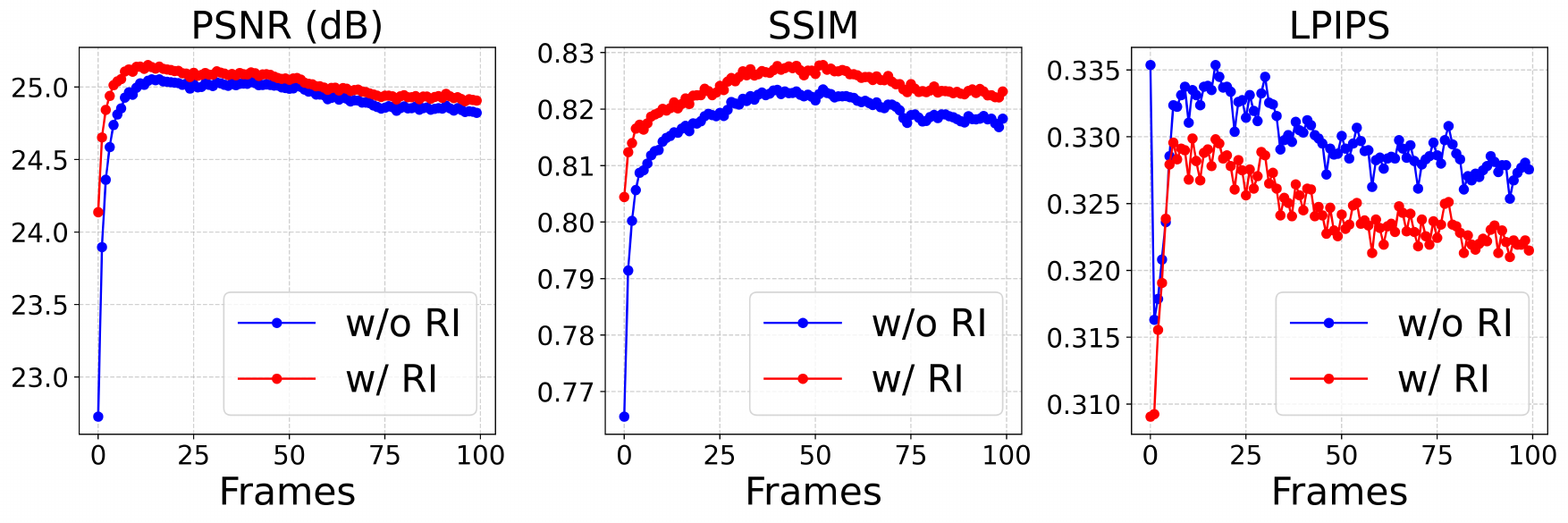}
    \caption{Ablation study of performance with and without reverse inference (RI).}
    \label{fig:reverse_ref}
\end{figure}

The quantitative results in the main paper have shown the superiority of incorporating reverse inference. Additionally, we statistically analyze the first 100 frames in the BVI-RLV \cite{Lin:BVIRLV:2024} test set. \cref{fig:reverse_ref} illustrates the comparison between the metrics of the online process and that of the offline process, which involves the reverse inference. The reverse inference approach achieves consistent improvements across all metrics, further proving our previous point of view.

\section{Discussion}
\label{sec:discussion}

Despite achieving relatively good results, our method still has several limitations, which also point out the direction for future research. Firstly, relying on explicit optical flow for frame alignment not only increases the computational overhead, but also the error in optical flow estimation affects the robustness of the method in dynamic scenes. Future work can explore implicit motion modeling or attention-based mechanisms to fuse temporal information more efficiently and stably without relying on explicit optical flow.

Secondly, our method assumes that the lighting between adjacent frames is relatively consistent, which may not hold true in scenes with rapidly changing lighting, such as flash and stroboscopic lighting. Extending this framework through transient lighting perception modeling to handle such challenging lighting variations will help enhance its practical applicability.

Furthermore, the current architecture only uses lightweight convolutional networks and does not introduce semantic guidance. Introducing semantic information  can achieve enhanced processing of content perception \cite{11092942}\cite{10205052}, especially in scenes with complex objects and textures, which is expected to improve the quality of the results. How to effectively combine semantic information with low-level video enhancement remains a direction worth exploring in future work.

\end{document}

%% file: img/histogram.tex
\begin{figure}[t]
    \centering
    \includegraphics[width=\linewidth]{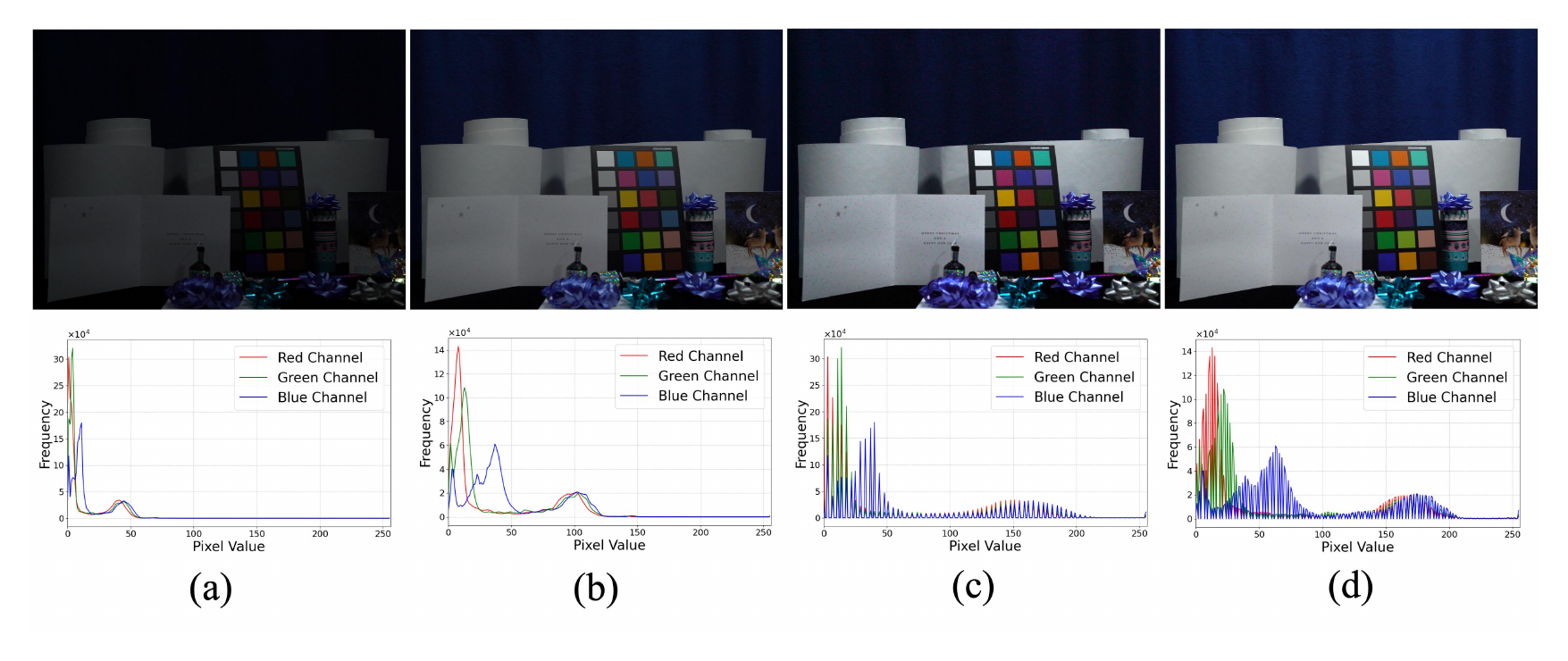}
    \caption{Histograms of inputs with 10\% and 20\% brightness ((a) and (b)) and corresponding BCP outputs $S_0$ ((c) and (d))}
    \label{fig:histogram}
\end{figure}

%% file: img/BVI_S17_clip.tex
\begin{figure*}[h]
  \centering
  \includegraphics[width=0.95\linewidth]{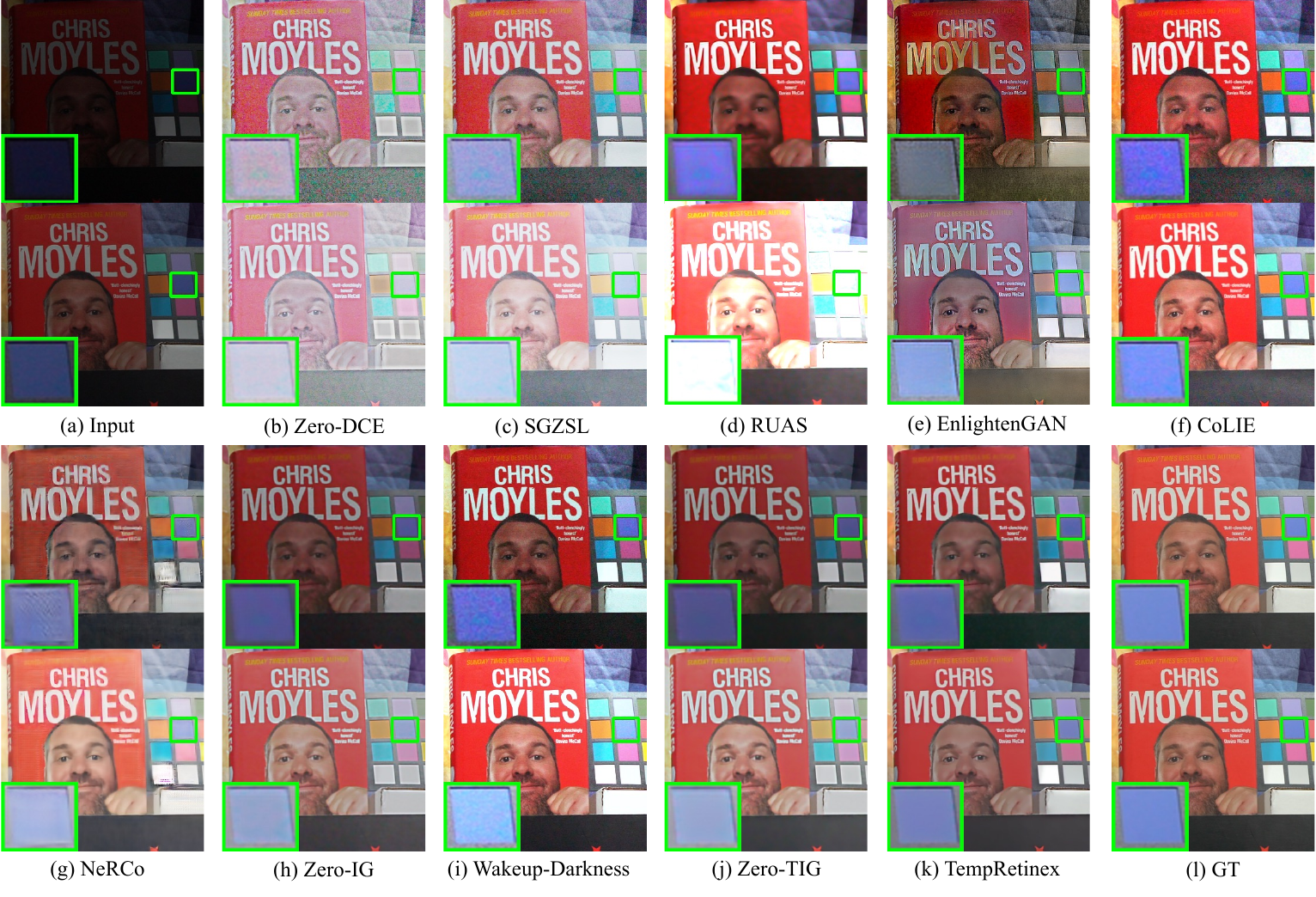}
  \caption{Visual comparison of unsupervised methods on the BVI-RLV under (top-row) 10\% and (bottom-row) 20\% brightness.}
  \label{fig:BVI_S17_clip}
\end{figure*}

%% file: img/DID_V51.tex
\begin{figure*}[h!]

  \centering
  \includegraphics[width=\linewidth]{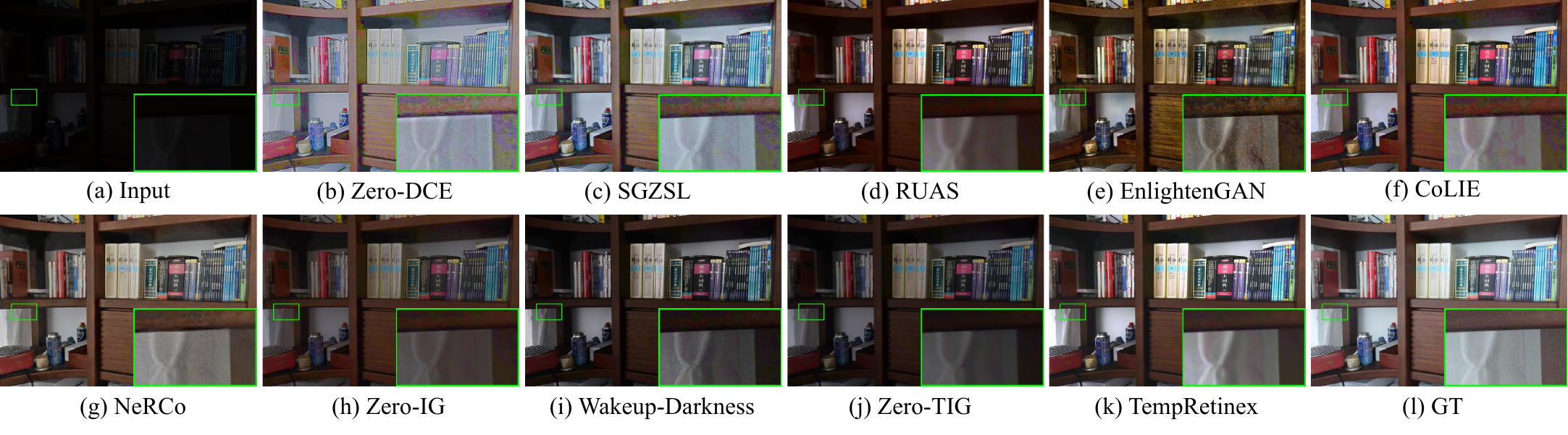}
  \caption{Visual comparison of unsupervised low-light enhancement methods on the DID.}
  \label{fig:DID_V51}
\end{figure*}

%% file: icme2025references.bib
@String(CVPR= {IEEE Conf. Comput. Vis. Pattern Recog.})

@String(ICCV= {Int. Conf. Comput. Vis.})

@String(ECCV= {Eur. Conf. Comput. Vis.})

@String(TIP  = {IEEE Trans. Image Process.})

@String(ICIP = {IEEE Int. Conf. Image Process.})

@String(AAAI = {AAAI})

@String(CVPR  = {CVPR})

@String(ICCV  = {ICCV})

@String(ECCV  = {ECCV})

@String(TIP   = {IEEE TIP})

@String(ICIP  = {ICIP})

@article{Lin:BVIRLV:2024,
  title={{BVI-RLV: A} Fully Registered Dataset and Benchmarks for Low-Light Video Enhancement},
  author={R. Lin and N. Anantrasirichai and G. Huang and J. Lin and Q. Sun and A. Malyugina and D. Bull},
  journal={arXiv:2401.10166},
  year={2024}
}

@inproceedings{huang:bvi:2025,
title = "BVI-Mamba: Video Enhancement Using a Visual State-Space Model for Low-Light and Underwater Environments",
author = "G. Huang and R. Lin and Y. Li and D. Bull and N. Anantrasirichai",
year = "2025",
booktitle = "Machine Learning from Challenging Data 2025",
}

@inproceedings{Lin:Low:2024,
author = {R. Lin and Q. Sun and N. Anantrasirichai},
title = {Low-light Video Enhancement with Conditional Diffusion Models and Wavelet Interscale Attentions},
year = {2024},
doi = {10.1145/3697294.3697304},
booktitle = {ACM SIGGRAPH CVMP}
}

@inproceedings{wang2021sdsd,
  title={Seeing Dynamic Scene in the Dark: High-Quality Video Dataset with Mechatronic Alignment},
  author={R. Wang and X. Xu and C. Fu and J. Lu and B. Yu and J. Jia},
  booktitle={ICCV},
  year={2021}
}

@inproceedings{Shi:zero:2024,
  title={{ZERO-IG: Zero}-Shot Illumination-Guided Joint Denoising and Adaptive Enhancement for Low-Light Images},
  author={Y. Shi and D. Liu and L. Zhang and Y. Tian and X. Xia and X. Fu},
  booktitle={CVPR},
  pages={3015--3024},
  year={2024}
}

@inproceedings{Zheng:semantic:2022,
  title={Semantic-guided zero-shot learning for low-light image/video enhancement},
  author={S. Zheng and G. Gupta},
  booktitle={IEEE/CVF WACVW},
  pages={581--590},
  year={2022}
}

@inproceedings{lin2024sunet,
  title     = {A Spatio-temporal Aligned SUNet Model for Low-light Video Enhancement},
  author    = {R. Lin and N. Anantrasirichai and A. Malyugina and D. Bull},
  booktitle = {IEEE ICIP},
  pages     = {1480--1486},
  year      = {2024},
  publisher = {}
}

@article{anantrasirichai2022artificial,
  title     = {Artificial Intelligence in the Creative Industries: A Review},
  author    = {N. Anantrasirichai and D. Bull},
  journal   = {Artificial Intelligence Review},
  volume    = {55},
  number    = {1},
  pages     = {589--656},
  year      = {2022},
  publisher = {},
  doi       = {10.1007/s10462-021-10036-6}
}

@article{jiang2021enlightengan,
  title={Enlightengan: Deep light enhancement without paired supervision},
  author={Y. Jiang and X. Gong and D. Liu and Y. Cheng and C. Fang and X. Shen and J. Yang and P. Zhou and Z. Wang},
  journal={TIP},
  volume={30},
  pages={2340--2349},
  year={2021},
  publisher={IEEE}
}

@inproceedings{Zero-DCE,
 author = {C. Guo and C. Li and J. Guo and C. Loy and J. Hou and S. Kwong and R. Cong},
 title = {Zero-reference deep curve estimation for low-light image enhancement},
 booktitle = {CVPR},
 pages    = {1780-1789},
 year = {2020}
}

@inproceedings{liu2021ruas,
title = {Retinex-inspired Unrolling with Cooperative Prior Architecture Search for Low-light Image Enhancement},
author = {R. Liu and L. Ma and J. Zhang and X. Fan and Z. Luo},
booktitle = {CVPR},
year = {2021}
}

@ARTICLE{10210621,
  author={X. Lv and S. Zhang and C. Wang and W. Zhang and H. Yao and Q. Huang},
  journal={IEEE TIP}, 
  title={Unsupervised Low-Light Video Enhancement With Spatial-Temporal Co-Attention Transformer}, 
  year={2023},
  volume={32},
  number={},
  pages={4701-4715},
  doi={10.1109/TIP.2023.3301332}}

@inproceedings{zhang2022towards,
  title={Towards unsupervised domain generalization},
  author={X. Zhang and L. Zhou and R. Xu and P. Cui and Z. Shen and H. Liu},
  booktitle={CVPR},
  pages={4910--4920},
  year={2022}
}

@inproceedings{narayanan2022challenges,
  title={On challenges in unsupervised domain generalization},
  author={V. Narayanan and A. Deshmukh and U. Dogan and V. Balasubramanian},
  booktitle={Workshop on Pre-registration in Machine Learning},
  pages={42--58},
  year={2022}
}

@inproceedings{li2025zerotig,
  title={Zero-TIG: Temporal Consistency-Aware Zero-Shot Illumination-Guided Low-light Video Enhancement},
  author={Y. Li and N. Anantrasirichai},
  booktitle={EUSIPCO},
  year={2025}
}

@inproceedings{ma2022toward,
  title={Toward Fast, Flexible, and Robust Low-Light Image Enhancement},
  author={L. Ma and T. Ma and R. Liu and X. Fan and Z. Luo},
  booktitle={CVPR},
  pages={5637--5646},
  year={2022}
}

@inproceedings{retinexformer,
  title={Retinexformer: One-stage Retinex-based Transformer for Low-light Image Enhancement},
  author={Y. Cai and H. Bian and J. Lin and H. Wang and R. Timofte and Y. Zhang},
  booktitle={ICCV},
  year={2023}
}

@inproceedings{teed2020raftrecurrentallpairsfield,
  title={Raft: Recurrent all-pairs field transforms for optical flow},
  author={Z. Teed and J. Deng},
  booktitle={ECCV},
  pages={402--419},
  year={2020},
  organization={Springer}
}

@inproceedings{shrivastava2023videodynamicspriorinternal,
author = {G. Shrivastava and S. Lim and A. Shrivastava},
title = {Video dynamics prior: an internal learning approach for robust video enhancements},
year = {2023},
booktitle = {NeurIPS},

}

@inproceedings{zhu2024unrolled,
  title={Unrolled Decomposed Unpaired Learning for Controllable Low-Light Video Enhancement},
  author={L. Zhu and W. Yang and B. Chen and H. Zhu and Z. Ni and Q. Mao and S. Wang},
  booktitle={ECCV},
  year={2024}
}

@inproceedings{mansour2023zero,
  title={Zero-shot noise2noise: Efficient image denoising without any data},
  author={Y. Mansour and R. Heckel},
  booktitle={CVPR},
  pages={14018--14027},
  year={2023}
}

@InProceedings{Fu_2023_ICCV,
    author    = {H. Fu and W. Zheng and X. Wang and J. Wang and H. Zhang and H. Ma},
    title     = {Dancing in the Dark: A Benchmark towards General Low-light Video Enhancement},
    booktitle = {ICCV},
    month     = {},
    year      = {2023},
    pages     = {12877-12886}
}

@inproceedings{Butler:ECCV:2012,
title = {A naturalistic open source movie for optical flow evaluation},
author = {D. Butler and J. Wulff and G. Stanley and M. Black},
booktitle = {ECCV},
pages = {611--625},
year = {2012}
}

@article{zhang2018perceptual,
  title={The Unreasonable Effectiveness of Deep Features as a Perceptual Metric},
  author={R. Zhang and P. Isola and A. Efros and E. Shechtman and O. Wang},
  journal={CVPR},
  year={2018},
  pages={586-595},
  url={https://api.semanticscholar.org/CorpusID:4766599}
}

@INPROCEEDINGS{Jiang:learn:2019,  
author={H. Jiang and Y. Zheng},  
booktitle={ICCV},   
title={Learning to See Moving Objects in the Dark},   
year={2019},  volume={},  number={},  pages={7323-7332},  
doi={10.1109/ICCV.2019.00742}}

@InProceedings{NerCO_2023_ICCV,
    author    = {S. Yang and M. Ding and Y. Wu and Z. Li and J. Zhang},
    title     = {Implicit Neural Representation for Cooperative Low-light Image Enhancement},
    booktitle = {ICCV},
    month     = {October},
    year      = {2023},
    pages     = {12918-12927}
}

@inbook{Colie2024,
  title = {Fast Context-Based Low-Light Image Enhancement via Neural Implicit Representations},
  booktitle = {ECCV},
  author = {T. Chobola and Y. Liu and H. Zhang and J. Schnabel and T. Peng},
  year = {2024},
  pages = {413–430}
}

@article{zhang2025Wakeup-Darkness,
  title={{Wakeup-Darkness}: When Multimodal Meets Unsupervised Low-light Image Enhancement},
  author={X. Zhang and Z. Xu and H. Tang and C. Gu and W. Chen},
  journal={TOMM},
  year={2025}
}

@inproceedings{lin2025towards,
  title={Towards a General-Purpose Zero-Shot Synthetic Low-Light Image and Video Pipeline},
  author={J. Lin and C. Morris and R. Lin and F. Zhang and D. Bull and N. Anantrasirichai},
  booktitle={McGE},
  pages={3--11},
  year={2025}
}

@InProceedings{zhang2021rethinking,
        author    = {Y. Zhang and H. Qin and X. Wang and H. Li},
        title     = {Rethinking Noise Synthesis and Modeling in Raw Denoising},
        booktitle = {ICCV},
        month     = {October},
        year      = {2021},
        pages     = {4593-4601}
    }

@inproceedings{wang2022exploring,
    author = {J. Wang and K. Chan and C. Loy},
    title = {Exploring CLIP for Assessing the Look and Feel of Images},
    booktitle = {AAAI},
    year = {2023}
}

@ARTICLE{NIQE,
  author={A. Mittal and R. Soundararajan and A. Bovik},
  journal={IEEE Signal Processing Letters}, 
  title={Making a “Completely Blind” Image Quality Analyzer}, 
  year={2013},
  volume={20},
  number={3},
  pages={209-212},
  doi={10.1109/LSP.2012.2227726}}

@INPROCEEDINGS{11092942,
  author={Gu, Yuxuan and Wang, Haoxuan and Ling, Pengyang and Wei, Zhixiang and Chen, Huaian and Jin, Yi and Chen, Enhong},
  booktitle={CVPR}, 
  title={Improving Visual and Downstream Performance of Low-Light Enhancer with Vision Foundation Models Collaboration}, 
  year={2025},
  volume={},
  number={},
  pages={16071-16080},
  doi={10.1109/CVPR52734.2025.01498}}

@INPROCEEDINGS{10205052,
  author={Wu, Yuhui and Pan, Chen and Wang, Guoqing and Yang, Yang and Wei, Jiwei and Li, Chongyi and Shen, Heng Tao},
  booktitle={CVPR}, 
  title={Learning Semantic-Aware Knowledge Guidance for Low-Light Image Enhancement}, 
  year={2023},
  volume={},
  number={},
  pages={1662-1671},
  doi={10.1109/CVPR52729.2023.00166}}
